\documentclass[%
twocolumn,
superscriptaddress,
 amsmath,
 amssymb,
 aps,
 prb,
 reprint,
]{revtex4-1}

\usepackage{graphicx}
\usepackage{bm}
\usepackage{braket}

\newcommand{\bb}{\beta}

\newcommand{\delm}{\delta_{\mu}}

\newcommand{\gm}{g_\mu}

\newcommand{\wcm}{\omega_{c\mu}}

\newcommand{\sig}[1]{\sigma_{#1}}
\newcommand{\sigk}[1]{\sigma_{#1,k}}

\newcommand{\gammc}{\gamma_{\mu c}}
\newcommand{\gammi}{\gamma_{\mu i}}

\newcommand{\Erion}[1]{\mbox{\ensuremath{^{#1}}Er\ensuremath{\mathrm{^{3+}}}}}
\newcommand{\YSO}{Y$_2$SiO$_5$}
\newcommand{\Er}[1]{\ensuremath{^{#1}}Er}

\newcommand{\NA}{{NA{}}} 
\renewcommand{\Er}[1]{\ensuremath{^{#1}}Er}
\renewcommand{\Erion}[1]{\mbox{\ensuremath{^{#1}}Er\ensuremath{\mathrm{^{3+}}}}}

\graphicspath{ {figures/} }

\begin{document}

\author{Gavin G. G. King}\affiliation{Dodd-Walls Centre for Photonic and Quantum Technologies, University of Otago, Dunedin, New Zealand}\affiliation{Department of Physics, University of Otago, Dunedin, New Zealand}
\author{Peter S. Barnett}\affiliation{Dodd-Walls Centre for Photonic and Quantum Technologies, University of Otago, Dunedin, New Zealand}\affiliation{Department of Physics, University of Otago, Dunedin, New Zealand}
\author{John G. Bartholomew}\altaffiliation{Current affiliation: 1. Centre for Engineered Quantum Systems, School of Physics, The University of Sydney, Sydney, NSW 2006, Australia; 2. The University of Sydney Nano Institute, The University of Sydney, NSW 2006, Australia.}\affiliation{Thomas J. Watson, Sr., Laboratory of Applied Physics, California Institute of Technology, Pasadena, CA, 91125, USA}
\author{Andrei Faraon}\affiliation{Thomas J. Watson, Sr., Laboratory of Applied Physics, California Institute of Technology, Pasadena, CA, 91125, USA}
\author{Jevon J. Longdell}\email{jevon.longdell@otago.ac.nz} \affiliation{Dodd-Walls Centre for Photonic and Quantum Technologies, University of Otago, Dunedin, New Zealand}\affiliation{Department of Physics, University of Otago, Dunedin, New Zealand}

\title {Probing Strong Coupling between a Microwave Cavity and a Spin Ensemble with Raman Heterodyne Spectroscopy}

\date{\today{}}

\begin{abstract}
    Raman heterodyne spectroscopy is a powerful tool for characterizing the energy and dynamics of spins. The technique uses an optical pump to transfer coherence from a spin transition to an optical transition where the coherent emission is more easily detected.
    Here Raman heterodyne spectroscopy is used to probe an isotopically purified ensemble of erbium dopants, in a yttrium orthosilicate (Y$_2$SiO$_5$)  crystal coupled to a microwave cavity. Because the erbium electron spin transition is strongly coupled to the microwave cavity, we observed Raman heterodyne signals at the resonant frequencies of the hybrid spin-cavity modes (polaritons) rather than the bare erbium spin transition frequency. Using the coupled system, we made saturation recovery measurements of the ground state spin relaxation time $T_1 = 10 \pm 3$ seconds, and also observed Raman heterodyne signals using an excited state spin transition. We discuss the implications of these results for efforts towards converting microwave quantum states to optical quantum states.
\end{abstract}

\maketitle

\section{Introduction}
\label{sec:intro}

Raman heterodyne spectroscopy\cite{Mlynek.1983,Wong.1983}  uses an optical pump beam to convert coherence on a spin transition to coherence on an optical transition, which generates an optical signal output. The optical photons have much higher energies than those in the radio frequency or microwave frequency regimes,  and hence are much easier to detect. Commonly, the optical pump beam is used as local oscillator, which allows the upconverted signal to be detected using heterodyne detection\cite{Yuen.1983}.
Raman heterodyne has proved very useful for characterizing the hyperfine structure of solid state dopants\cite{ Manson.1990,Ma.2018, Xiao.2020} especially those, like the rare earths, with narrow optical transitions. This characterization and the ability  to measure weakly doped samples has allowed extremely long coherence times to be observed\cite{Zhong.2015, Ortu.2018}.

Raman heterodyne processes also hold promise for efficient conversion of microwave to optical frequency signals for quantum information applications\cite{Lauk.2020}. This conversion is desirable because optical photons can be distributed long distances in low loss fiber networks and couple more strongly to atoms and atom-like systems.  Furthermore, optical frequency modes at room temperature are not populated, avoiding thermal noise. In contrast, single microwave frequency excitations, which couple naturally to superconducting qubits, are swamped by thermal noise at all but milli-Kelvin scale temperatures. A diverse range of approaches are being investigated for microwave to optical conversion\cite{Lambert.2020,Lauk.2020}. The highest efficiencies ($\approx 50\%$) to date\cite{Higginbotham.2018}  have been demonstrated using a mechanical oscillator made from a thin crystalline membrane. However the conversion was narrow bandwidth and not completely free from noise.
Recently higher bandwidth, lower efficiency, upconversion using superconducting qubits as the source of the microwave photons has been reported using nanomechanical resonators\cite{Fang.2016, Bagci.2014,Mirhosseini.2020}. Electro-optic approaches\cite{Rueda.2019, Zhang.2019}, and atomic systems\cite{Gard.2017, Covey.2019, Kiffner.2016, Han.2018, Vogt.2019} as well as color centers in crystals\cite{Marcos.2010} are also actively being pursued.

Rare earth ion dopants are another exciting approach\cite{Thiel.2011,Williamson.2014,OBrien.2014,Welinski.2019, Miyazono.2016,OBrien.2016,Probst.2014b}. The rare earths are characterized by their narrow optical and spin transition linewidths, especially at low temperatures. 
Specifically, these narrow inhomogeneous transitions ($\sim$100\,MHz optical and $\sim$10\,MHz microwave) can occur despite high ion concentration ($\sim$100\,ppm), which results in very high spectral densities. 
Given that the efficiency of the rare earth ion transduction schemes\cite{Williamson.2014, OBrien.2014} depend on the achievable spectral density rather than simply the ensemble size, it is possible to achieve high number conversion efficiency ($\sim$10\%) in both bulk\cite{fernandez-gonzalvo.2015, Fernandez-Gonzalvo.2019} and on-chip miniaturized architectures\cite{Bartholomew.2020}. The rare earth ion platform also has the potential to operate at even higher rare earth concentration using stoichiometric, as opposed to doped, materials\cite{Everts.2019, Everts.2020}, at telecommunication wavelengths, at zero applied magnetic field\cite{Welinski.2019, Rakonjac.2020, Bartholomew.2020}, and with incorporated quantum state storage\cite{OBrien.2014}, making it an appealing system for hybrid quantum technology beyond direct transduction.

Previously, conversion of 4.7\,GHz photons to 1536\,nm photons has been carried out using a combined microwave and optical resonator containing an erbium-doped yttrium orthosilicate (\Er{}:\YSO{} or \Er{}:YSO) crystal\cite{fernandez-gonzalvo.2015,Fernandez-Gonzalvo.2019}. The upconversion process involved three connected transitions in the \Erion{} ion, as shown in Fig.~\ref{fig:config}d and Fig.~\ref{fig:config}e.
The efficiency there was limited to $10^{-5}$ by two factors: optical absorption from the unwanted \Er{167} isotope, and the relatively high temperature at which the experiments were carried out.
Unlike the even isotopes of erbium, \Er{167} (28.6\% natural abundance) has hyperfine structure. This spans about 5\,GHz for both the $Z_1$ and $Y_1$ states, and results in a complicated, broadband, poorly resolved hyperfine structure on the  $Z_1\rightarrow Y_1$  transition\cite{Rakonjac.2020}. Temperature is important because the conversion efficiency is proportional to the square of the population difference between the two spin states. The population difference is only 1 part in 17 at 4\,K for a 4.7\,GHz transition.

Here we explore upconversion processes in a crystal with isotopically purified \Er{170} dopants at low temperatures ($<$1\,K) using a single optical pass through the sample.  Models suggest that with an optical resonator it is ultimately possible to achieve efficiencies of around 80\% in this regime \cite{Barnett.2020}. Our results are compared with these models. It also is the first time to our knowledge where Raman heterodyne spectroscopy has been used when the spins being probed are strongly coupled to a microwave resonator. We also investigate the spin lattice relaxation times.  While a microwave-optical converter for quantum information applications will only need to deal with very weak input signals, the long `reset time' caused by a long spin-lattice relaxation time $T_1$ could be a practical issue for microwave-optical conversion at the quantum level. 

\section{Configuration and Methods}
\label{sec:config}

\begin{figure}
    \includegraphics[width=\columnwidth]{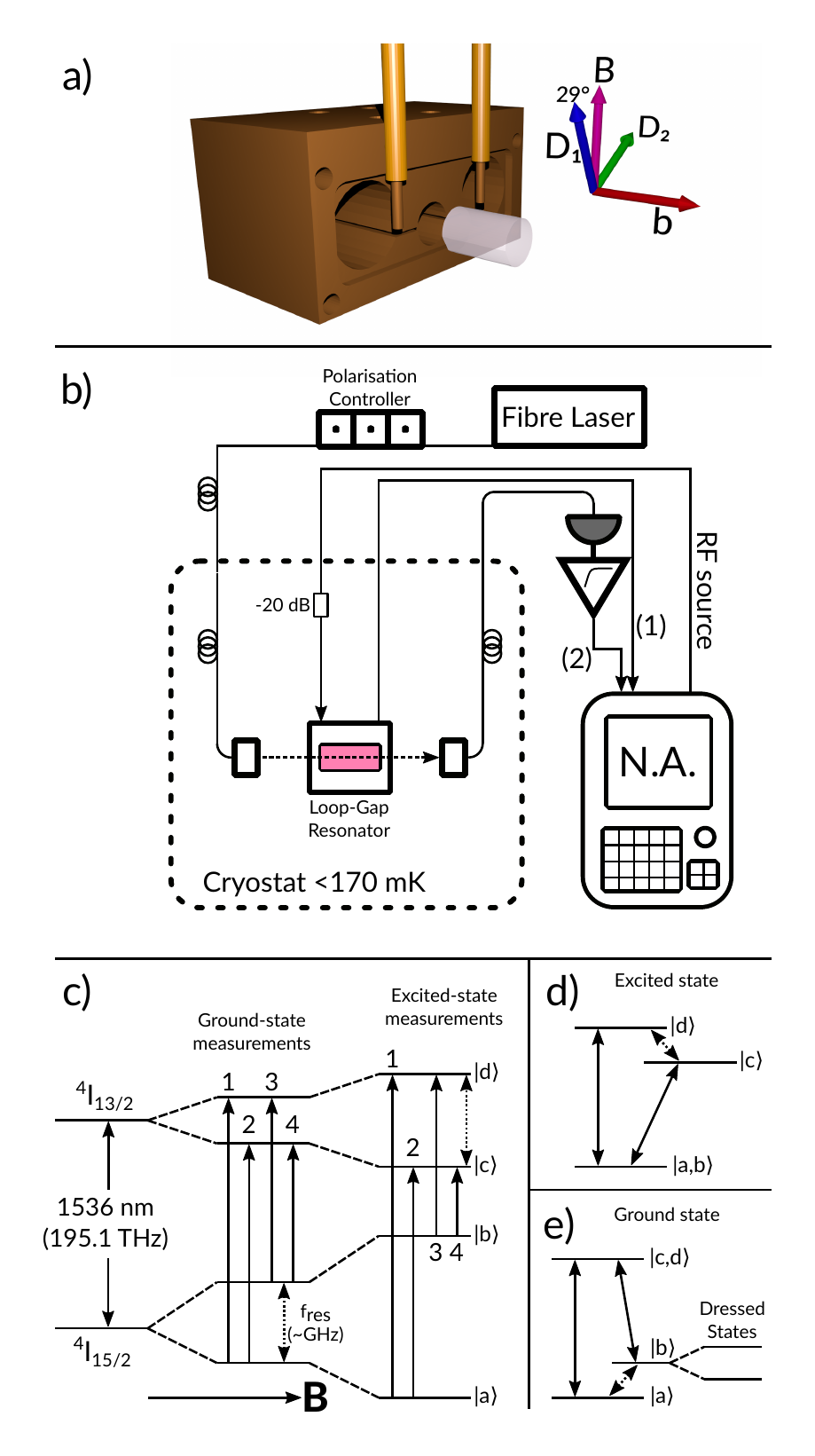}
    
    \caption{\label{fig:config}(Color online) a) The loop loop-gap resonator, with the sample removed from the center of the cavity. \textbf{B} is the applied magnetic field; $b$, $D_1$, and $D_2$ are the axes of the \Erion{}:YSO sample. The light propagates along $b$, which is also the direction of the RF magnetic field in the sample. The ends of the resonator have been removed for clarity.
b) A schematic of the apparatus. Cable (1) was connected to the \NA{} input to measure the  microwave transmission. Cable (2) was connected when measuring the Raman heterodyne signal. 
c) A schematic of the relevant energy levels in the \Erion{} ion, as described in the text. 
The arrows show the four possible optical pump transitions, which we label 1 to 4. The dashed arrow indicates the transition resonant with the microwave resonator in each case.
d) and e) Different arrangements for generating Raman heterodyne measurements, which are discussed in the text. 
    }
\end{figure}

The crystal of erbium doped \YSO{} used was cut from a custom-grown boule, doped at 50\,ppm with \Erion{170} substituting for the Y$^{3+}$ ions. The boule was grown by Scientific Materials, Bozeman, Montana, from precursor materials with 97\% isotopically pure \Er{170}. 
Of the isotopes with no nuclear spin, \Er{170} was selected, because in the catalog of the isotopes available, \Er{170} had the smallest amount of \Er{167} impurities. This is despite \Er{170} having lower natural abundance (14.93\%) than both \Er{168} (26.78\%) and \Er{166} (33.61\%).
The sample was a cylinder of 5\,mm diameter and 8\,mm length. The crystal $b$ axis was along the cylindrical axis. The two ends were polished but uncoated. A small reference flat was cut into the curved side parallel to indicate the crystal $D_1$ axis.

The experimental apparatus comprised the crystal of Er:\YSO{} inside a  5\,GHz microwave loop-gap resonator, as shown in Fig.~\ref{fig:config}a. Around the microwave resonator  were a set of Helmholtz coils, and a pair of fiber-optic collimators to direct a laser beam through the sample. This whole assembly was mounted on the mixing chamber stage of a dilution refrigerator.
 The design of the resonator is described elsewhere\cite{fernandez-gonzalvo.2015,Fernandez.Gonzalvo.2017}. A pair of semi-rigid coax lines (silver-plated copper, 2.197\,mm OD, UT-085) were stripped to make a pair of electric dipole antennae, which were located to couple into the electric field mode of the resonator. 

The magnet assembly comprised a pair of Helmholtz coils to provide a magnetic field perpendicular to the $b$  axis of the sample. A second smaller pair of Helmholtz coils provided a field along the $b$ axis, to allow the field to be tuned to be exactly in the $D_1 -D_2$ plane. The overall direction of the field relative to the sample is shown in Fig.~\ref{fig:config}a. These coils were home-made from NbTi superconducting wire (Supercon Limited, SC-T48B-M-0.3mm).

We addressed the erbium ions at site 1 in YSO, where the optical transition is approximately 1536.4753\,nm. Site 1 has two magnetically inequivalent sub-classes related by a $C_2$ rotation about the $b$ axis. These sub-classes have degenerate $g$-tensors  when the magnetic field is in the $D_1-D_2$ plane, or along the $b$ axis.\cite{Sun.2008}

We operated in the first of these degenerate regimes, with the field orientated at an angle of $\theta\approx29^\circ$ from $D_1$ in the direction of $D_2$.  
This angle was chosen to maximize the magnetic dipole transition strength for the upper-to-lower (like-to-unlike, or transitions 1 and 4 in  Fig.~\ref{fig:config}c) transitions between Zeeman split levels in the excited and ground states. 

A fiber laser\footnote{Koheras Adjustik K82-151-12.} provided light in a single mode fiber. Inside the dilution fridge, the fiber output was collimated with a free-space lens in a titanium housing\footnote{Model 60SMS-1-4-A2.7-45-Ti, Sch{{\"{a}}}fter + Kirchhoff GmbH, Kieler Stra\ss{}e 212, D-22525 Hamburg} and transmitted through the crystal.
The free-space beam passed along the $b$-axis of the sample and coupled back into optical fiber with an identical collimator. Outside the cryostat the light was detected using either a standard InGaAs or a high speed \footnote{Hamamatzu G7096, InGaAs, 10\,GHz} photodetector.

The apparatus was designed to minimize magnetic field inhomogeneity across the sample volume, by using oxygen-free copper throughout.
Commercial titanium amagnetic collimators were used, because copper collimators were not available. While titanium is a  superconductor, the critical field to suppress superconductivity is very low and we assumed that the effect was negligible. 
Furthermore,  Helmholtz coils have a null or line of zero field  on a circle coaxial with the coils, centered between them. To minimize the perturbation of the magnetic field at the samples, the centers of the  collimators were mounted as close to this null as possible.

It was not possible to adjust the collimators once the cryostat was closed. The collimators were carefully selected for their stability, and the mounts carefully designed to be as symmetric about the optic axis as possible.
As the system was cooled from room temperature to the base temperature (25\,mK), the fiber to fiber transmission decreased from 60\% to 22\%. 

A three level system can be formed from any three of the four energy levels of the two Kramers doublets, as shown in Figs.~\ref{fig:config}c, d, e.
As the magnetic field \textbf{B} is increased, each of the two Kramers doublets split into two. 
We measured the ground state spin transitions with the loop-gap resonator in resonance with the $Z_1$ doublet, and the excited state spin transitions with the resonator in resonance with the $Y_1$ doublet, as shown by the dashed arrows in Fig.~\ref{fig:config}c.
The solid arrows in Fig.~\ref{fig:config}c show the four optical pump transitions which we label 1 to 4.

Three types of measurements were performed: optical absorption, microwave cavity transmission, and Raman heterodyne detection.

To measure optical absorption spectra the fiber laser frequency was set at one edge of the spectrum  and allowed to settle. The laser frequency set-point was then set to the other edge and the transmission monitored for the 8--12 seconds it took for the laser to tune to the new frequency. 
To speed up data collection spectra were taken alternately sweeping up and then sweeping down in laser frequency.

Microwave cavity transmission was used to measure the interaction between the spins and the cavity. The magnetic field was stepped, and the transmission through the cavity was measured with a network analyzer (NA)\footnote{Agilent Fieldfox N9912A.} for each field.

Raman heterodyne detection was used to measure the conversion between microwave photons and optical photons. 
For the Raman heterodyne measurements the microwave (RF) signal was sent to the resonator with the same input antenna as for the cavity transmission measurements. 
The RF signal from the high-speed photodetector was then amplified and filtered in a band between 4.0 and 5.5\,GHz, before being measured by the \NA. 
A two-dimensional map was made by first stepping the laser, then for each laser frequency stepping the magnetic field over the desired range, and the recording a spectrum with the \NA{} field and laser frequency. To speed data collection, the field was stepped first up then down as the laser was stepped.

To gain further insight into the process, and to verify our models of the system, the results were simulated using the input-output formalism\cite{Gardiner.1985}. We simulated both the microwave transmission and the Raman heterodyne processes, and the results of these simulations are plotted next to the measured data. The method used is briefly detailed in appendix~\ref{sec:simulate}, and in much more detail in Ref.~\onlinecite{Barnett.2020}. 

Throughout these measurements, the dilution fridge was run at it lowest possible temperature (for the given heat load). Our main means of measuring the temperature was the temperature sensing resistor mounted on the mixing chamber plate, with our apparatus attached to this plate. The temperature recorded was between 40\,mK and 180\,mK during the below measurements. 
Furthermore, the thermometer was mounted on the coldest plate of the cryostat, right next to the mixing chamber, while our apparatus was mounted off the plate, attached through several different pieces of metal. Because of this thermal distance, it is likely there was a thermal gradient between the sample and the thermometer.
It is also possible to infer the temperature from the data  which we will discuss with the results. 

\section{The Electronic Ground State}
\label{sec:ground}


\begin{figure}
    \includegraphics{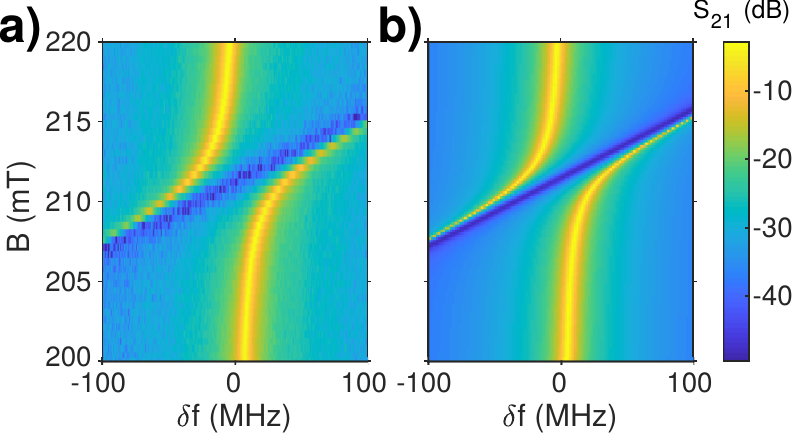}
\caption{\label{fig:avoidedcrossing:epr}(Color online) a) Microwave cavity transmission as a function of magnetic field strength. b) Simulation of a). Strong coupling between the erbium spins results in an avoided crossing as the spins are tuned through the microwave resonator.  In this figure, $\delta{}f=0$ corresponds to 5020\,MHz. The color scale shows the overall transmission efficiency from one \NA{} port to the other.
    }
\end{figure}

\begin{figure}
    \includegraphics{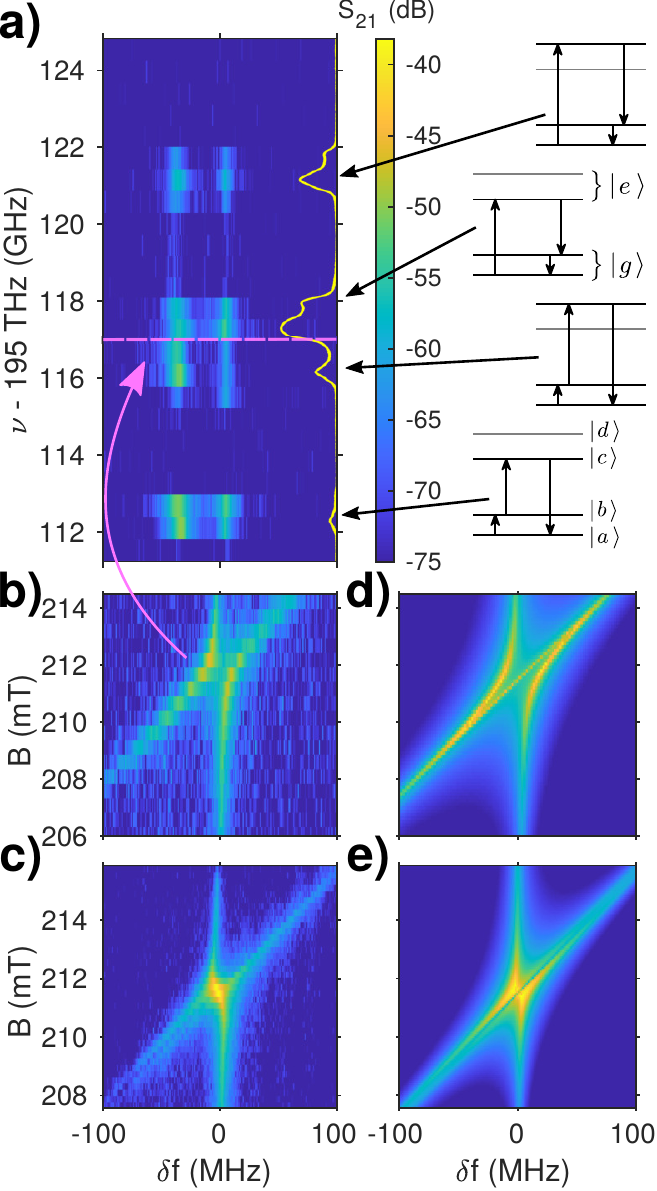}

    \caption{\label{fig:groundstate}(Color online) Raman heterodyne signal measured in the ground state. 
a) Raman heterodyne scan at a fixed magnetic field, with optical absorption spectrum (yellow). The horizontal dashed line indicates the optical frequency for parts b) and c) were collected.
 For part b) the microwave power was -15\,dBm at entry to the fridge, a power level chosen to limit the saturation of the spins. For part c) the microwave power was 8\,dB higher.
d) and e) show simulations of b) and c) respectively.
In all these figures, the bare cavity frequency is 5020\,MHz. The pump laser frequency is denoted by  $\nu$.
The color scale is common to all these figures, and shows the overall efficiency from one \NA{} port to the next.
}
\end{figure}

To observe Raman heterodyne produced by a ground state coherence we tuned the magnetic field to make the cavity near resonant with the  transition between the two levels of the electronic ground state ($^4 I_{15/2} Z_1$) doublet. The calculated\cite{Sun.2008} $g$-factor for the RF magnetic field along the $b$ crystallographic axis is 8.84.

Microwave cavity transmission and the Raman heterodyne signal were recorded as a function of applied microwave frequency, optical frequency and magnetic field. The optical absorption spectrum was measured as a function of magnetic field, from which we measured the effective $g$-factor. For the ground levels $\ket{a}\leftrightarrow\ket{b}$ the $g$-factor was 1.72, and for the excited levels $\ket{c}\leftrightarrow\ket{d}$ it was 1.28.

The microwave cavity transmission is shown in Fig.~\ref{fig:avoidedcrossing:epr}a, using the experimental configuration shown in Fig.~\ref{fig:config}b. Fitting a Gaussian curve to the microwave transmission spectrum at a magnetic field such that the ions were greatly detuned from the resonator, we measured the microwave transition (spin) inhomogeneous linewidth as 3\,MHz.  
A strong avoided crossing is visible between the vertical cavity resonance and the diagonal resonance in the \Erion{} ions, which is a characteristic of strong coupling\cite{Tabuchi.2014,Probst.2014b}.
We determined the coupling strength between the \Erion{} spins and the microwave resonator by measuring the splitting between the two modes in the avoided crossing.
The mode splitting was $74\pm1$\,MHz, giving a coupling strength $\sqrt{N}g = 37\pm0.5$\,MHz. 
The temperature measured at the mixing chamber was 55--65\,mK. In the model, to achieve the correct size of anticrossing the temperature was set to 150\,mK. Part of the reason for this discrepancy could have been the fact that we were unsure of the precise concentration of the erbium ions\cite{Bottger.2006}.

Note that there is a null in the transmission along a diagonal line following the resonance of the ions. This is because along this line the two ion-cavity dressed states are excited in such a way as their cavity components cancel.

The Raman heterodyne results for the ground state are shown in Fig.~\ref{fig:groundstate}. 
Fig.~\ref{fig:groundstate}a shows the Raman heterodyne signal as a function of applied microwave frequency and applied laser frequency with the magnetic field tuned to the anticrossing. 
There are two columns of spots where Raman heterodyne signal is seen, corresponding to the ion--cavity dressed states (Fig.~\ref{fig:config}e), rather than to the microwave spin transition in erbium.
Fig.~\ref{fig:groundstate}a also shows a number of rows of spots corresponding to the optical transitions. The optical absorption spectrum is shown along the $y$-axis.  

If the temperature was sufficiently low, there would be no thermal population in the state $\ket{b}$, and we would only see two optical absorption lines originating from the $\ket{a}$ level. Because we are operating at around 50\,mK to 150\,mK the thermal population of state $\ket{b}$ is around 10\% of the population of state $\ket{a}$. We thus expect four optical absorption peaks corresponding to the four transitions indicated.  
However each of these optical transitions doesn't appear as a single peak, instead there is slightly resolved structure. The origin of the structure is unclear, but it is potentially due to inhomogeneous strain inside the crystal. 
Between the excited state measurements and the ground state measurements we reamed out the central hole of the resonator to reduce strain on the sample as the resonator cooled. The amount of structure on the absorption lines was thus reduced in these later ground state measurements. 
It is difficult to define a linewidth given this structure. However, in order to give an indication of the inhomogeneous linewidth, we fitted a single Gaussian to each of the four transitions. For each of the four transitions 1--4 in Fig.~\ref{fig:config}c, the Gaussian linewidth $\sigma$ is approximately $\sigma_1=270$\,MHz, $\sigma_2=410$\,MHz, $\sigma_3=150$\,MHz, and $\sigma_4=200$\,MHz.
 
For these ground state measurements the optical pump power was around 2\,mW at the input to the crystal. 
The RF signal was $-15$\,dBm at the entrance to the cryostat. After further attenuation by the cables and attenuator, we estimate the power at the input to the resonator was around $-40$\,dBm. 

The detected signal is largest for the optical transitions from the upper ground state $\ket{b}$, being approximately 20\,dB stronger than the signals from the ground state $\ket{a}$. In contrast the optical absorption is strongest for the transition from the lowest state $\ket{a}$. Both of these effects can be explained by the population difference between $\ket{a}$ and $\ket{b}$. The amount of absorption is directly proportional to the population in the lower level, whereas for Raman heterodyne it is more favorable to have the laser addressing an optical transition where the lower level is empty.
 This is potentially because the upper state $\ket{b}$ has much less thermal population than the lowest state $\ket{a}$, so that there are fewer ions to absorb light for the optical absorption, whereas for the Raman heterodyne there are fewer ions to be driven down from $\ket{b}$ to $\ket{a}$, thus increasing the signal seen. 
Optical absorption raised the recorded temperature of the mixing chamber to between  between 130\,mK and 150\,mK. The temperature of the sample is likely to be higher, as described when we compare the Raman heterodyne measurements and models.

For a fixed optical frequency, around 195.117\,THz (see Fig.~\ref{fig:groundstate}a), we investigated the effect of microwave power on the Raman heterodyne signal. This is shown in Fig.~\ref{fig:groundstate}b and Fig.~\ref{fig:groundstate}c.  
With the lower microwave power, Fig.~\ref{fig:groundstate}b, $-15$\,dBm (corresponding to around -40\,dBm input to the resonator) we see there is a clear avoided crossing. 
The separation is around 28\,MHz, so the coupling strength is around 14\,MHz, less than the 74\,MHz seen in Fig.~\ref{fig:avoidedcrossing:epr}. This is attributed to the temperature being higher for these measurements, because in addition to the microwaves there is now light heating the sample. Using the model, we simulated this measurement, shown in Fig.~\ref{fig:groundstate}e, and found the results are consistent with a temperature of around 670\,mK, rather than the recorded temperature of 150--160\,mK. 

The model also shows a diagonal bright line between the two dressed states, following the bare-spin resonant frequencies, and hints of this are visible in the measured data. We call this behavior an ``avoided-avoided-crossing''. It is a nonlinear effect, only occurring for large driving strengths when saturation of the spins causes the strong coupling effects to disappear. The same behavior has also been observed in the interaction between rare earths and optical resonators.\cite{Fernandez.Gonzalvo.2017}

Increasing the microwave power by 8\,dBm (Fig.~\ref{fig:groundstate}c), caused the strength of the signal to increase, but also caused the avoided crossing to disappear. This indicates either saturation of the spins or an increase in the crystal lattice temperature, or both. 
The recorded mixing chamber temperature for these high microwave power measurements was around 180\,mK, but the model indicates the spin temperature was closer to 1.7\,K. 

In the simulated data (Fig.~\ref{fig:groundstate}e) there is a clear dark line through the center of the crossing. This dark line is just resolvable in the measured data (Fig.~\ref{fig:groundstate}c) and  also seen in the excited state (Fig.~\ref{fig:excitedstate}). \label{sec:darkline} 
This dark line shows reduced Raman heterodyne signal when the microwave Rabi frequency is large enough to start to saturate the ions' homogeneous linewidth and when the microwave drive frequency is precisely in the center of the inhomogeneous line. The Raman heterodyne signal can be broken up in to two parts: the signal from the ions that have a transition on resonance with the laser, and those that are detuned. When in the center of the inhomogeneous line and at high driving powers both of these two contributions is small. The signal from the resonant ions is small due to saturation, the signal due to the off resonant ions is small because the contribution from  the positively and negatively detuned ions cancel.

\section{The Electronic Excited State}
\label{sec:excited}


\begin{figure}
    \includegraphics{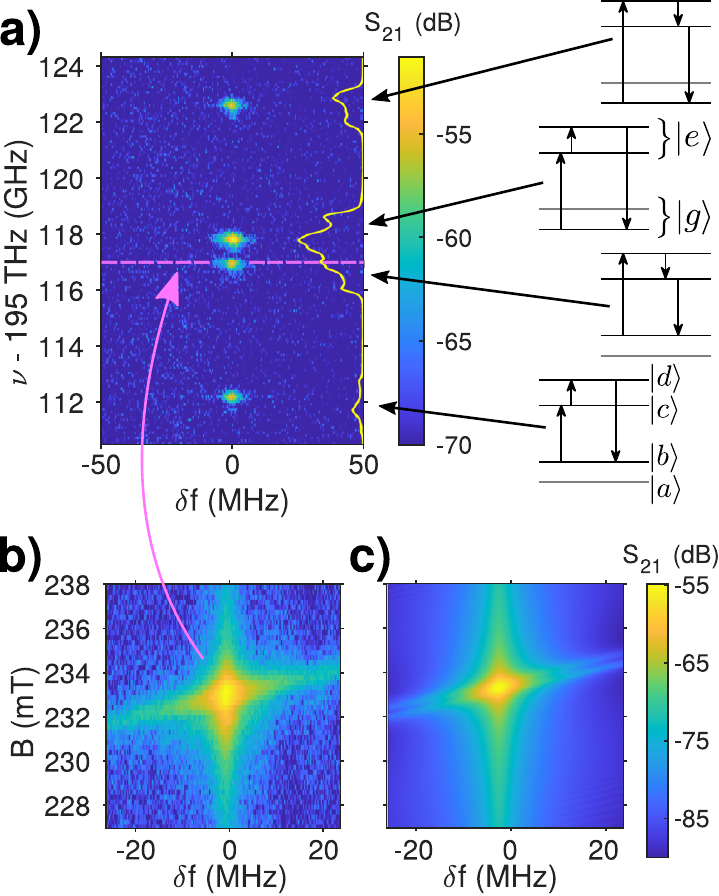}
    
    \caption{\label{fig:excitedstate}(Color online) a) Raman heterodyne signal in the excited state. 
    An optical absorption spectrum is plotted against the right-hand axis.
    The four spots correspond to the four transitions as shown. The horizontal dashed line indicates the optical frequency where the data in b) was collected.
    The color bar scale is in dB transmitted from one port of the \NA{} to the other, and is common to all these sub-figures.
    b) Changing the magnetic field, the microwave resonator frequency (vertical) doesn't change, while the frequency of the ions does.
    c) Simulated version of b). 
    The empty cavity frequency is 4732\,MHz in these figures.
}

\end{figure}

To observe Raman heterodyne produced by an excited state coherence we tuned the magnetic field to make the cavity near resonant with the  transition between the two levels of the electronic excited  state ($^4 I_{13/2} Y_1$) doublet. The effective $g$-factor along the $b$-axis was calculated as 10.0 using the spin Hamiltonian from Ref.~\cite{Sun.2008}.

In Fig.~\ref{fig:excitedstate}a, we see the Raman heterodyne signal has four spots corresponding to the four expected optical transitions. In this excited state case there is only a single column of spots, because there is no strong coupling between the cavity and ions to cause dressed states.

The optical absorption spectrum is shown on the right-hand axis of Fig.~\ref{fig:excitedstate}a. 
We see the four optical transitions shown in the energy level diagrams because there is thermal population in the $\ket{b}$ level, as explained in the previous section.
Like Fig.~\ref{fig:groundstate}a, each of the optical transitions has a broad peak with poorly resolved structure. Interestingly, such broad optical peaks aren't seen in the Raman heterodyne spectrum, where four well resolved spots can be seen. This suggests a strong correlation between optical and spin frequencies which we will discuss later. Measuring the optical spectrum as a function of applied magnetic field, we found that the $g$-factor for the ground state levels $\ket{a}$ and $\ket{b}$ was 1.50, and for the excited state levels $\ket{c}$ and $\ket{d}$ was 1.45. These differ from the values in the ``Ground State'' measurements, because the sample was removed and replaced in the resonator between these measurements and those. Because the structure on the lines here is much more significant than that in Fig.~\ref{fig:groundstate}a, we did not fit a curve to determine the inhomogeneous linewidth. In any case, it is very difficult to separate the two central peaks from one another.

Fixing the laser frequency to 195.117\,THz, we investigated the effect of changing the magnetic field by stepping the magnetic field and recording a Raman heterodyne spectrum  for each field, as shown in Fig.~\ref{fig:excitedstate}b. The vertical cavity mode and the diagonal electron spin resonance are visible, with a peak where they intercept. 
Because the majority of the atoms are not in the excited state but the ground state, the ensemble of ions does not strongly couple to the microwave resonator, and so we do not see an avoided crossing.

These results were modeled, as shown in Fig.~\ref{fig:excitedstate}c. 
The estimated temperature from the model was 150\,mK, which is close to the measured temperature of 120--130\,mK. 
In the model we see a diagonal dark line. The mechanism for this is the same as for the ground and is discussed at the end of Sec.~\ref{sec:darkline}.

\section{The Spin-Lattice Relaxation Time $T_1$}
\label{sec:relax}
\begin{figure}
    \includegraphics[width=\columnwidth]{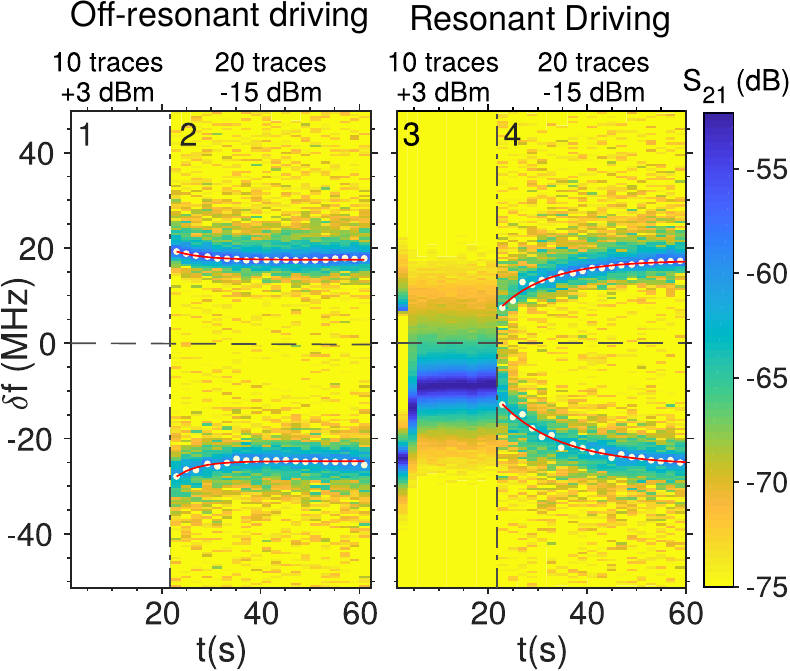}
    
    \caption{\label{fig:saturationthumping}(Color online)  Measurements of the $T_1$ time of the system by saturation recovery. 
        The dots and lines relate to the fit as described in the text.
        Part 1 is empty because the resonator was driven at a different frequency, as explained in the text.
        The horizontal dashed line is the resonance frequency of the empty cavity. The vertical dot-dash lines separate the different powers and frequency ranges.
        The power shown is the output of the \NA{}, and $\delta$f=0 corresponds to 5020\,MHz.
}
\end{figure}


The spin-lattice relaxation time $T_1$ was measured by saturating the spins, and measuring how long it takes for them to return to thermal equilibrium. While they are relaxing, the population difference can be monitored by observing the size of the avoided crossing.  This was measured with the same physical arrangement as for the microwave transmission, but with the magnetic field fixed at close to the center of the avoided crossing, while changing the microwave power. 
The results are shown in Fig.~\ref{fig:saturationthumping}. 

The measurement consisted of four parts. First, a strong microwave signal was applied with the microwave frequency sweep centered on a different mode (2149\,MHz) of the microwave resonator. Second, a weak signal was used to probe the size of the avoided crossing for approximately 40\,s. For the third part, the microwave power of this `probing' signal was increased to a point where it strongly drives the ions. It was left at this power level for about 20\,s. Finally the network analyzer power turned to the same low power settings used in step two, and the size of the anticrossing was monitored as the population difference returned.

The purpose of the first two steps is to rule out heating of the sample as a reason for the reduced population difference. By driving a different mode of the resonator with the same power, we heat the resonator the same amount. We see no change from this heating. However, driving away from the spin resonance seems to increase the population difference, because when the low-power scan is resumed the avoided crossing reduces in size for about 20\,s before reaching a steady state. This is because the spins aren't being driven by the weak probing field while the strong off-resonant field is applied.

In Fig.~\ref{fig:saturationthumping}, the white dots show the peaks of transmission. The red lines are fitted to these peaks. An exponential curve is fitted to the difference between the red lines to give the time constant {$T_1$}. 
The splitting is proportional to $\sqrt{N}$, where $N$ is the population difference between the upper $\ket{b}$ and lower $\ket{a}$ states. There is a transition rate between the states with a fixed relationship according to the Boltzmann distribution. The dominant effect is the $\ket{b}\rightarrow{}\ket{a}$ rate $1/T_1$, such that the splitting $\delta f$ is given as: 
\begin{equation}
  \delta f \propto \sqrt{N} \propto \sqrt{1-\exp{\left(\frac{-t}{T_1}\right)}}
\end{equation}

The measured time constant in the figure shown was  $10 \pm 2\,$s. Over several sets of data, collected with different saturating times and powers, it varied within $10\pm3\,$s. 

\section{Discussion}
\label{sec:discuss}
We observed Raman heterodyne signals when both using the ground and excited state spin transitions, demonstrating the ability to do microwave optical conversion with both. 
A number of advantages have been suggested for using excited state spin transitions for upconversion\cite{Welinski.2019}. There is significantly less cross relaxation between the spins because the effective concentration of the dopants is very low. There is also less trouble with ``parasitic'' spins which are those which due to their location can absorb microwave photons but never interact with the optical fields. Our results show that another benefit is that the microwave cavity frequency doesn't suffer frequency pulling from the ions. A trade off for this is that now both optical transitions will have significant population in their ground states, meaning that the frequencies of both optical resonators modes will be pushed around by the ions.

Raman heterodyne spectroscopy has a long history of providing useful information on the spectroscopy of spins. However, in the ground state spin transitions the Raman heterodyne signals occurred when the microwave drive was on resonance with the ion-cavity dressed states rather than  when on resonance with the ions, thus showing the utility of Raman heterodyne for also  probing spin-cavity coupling.

There is a lot of structure on the optical spectra, even after attempting to reduce compression strain by increasing the bore of the resonator. The structure is not present in the Raman heterodyne spectra. This is clearly so in the excited state (Fig.~\ref{fig:excitedstate}a). In the ground state data (Fig.~\ref{fig:groundstate}a) the resolution is lower, but careful consideration of the data, including ``slices'' from higher and lower magnetic fields, shows that if  the structure is present at all, it is much smaller than would be expected from the optical spectrum. 
One explanation is that there is a strong correlation between the optical frequencies and the spin frequencies, which means that the microwave cavity, in selecting one spin transition frequency, has also selected out an ensemble of ions with a narrow optical spectrum. Such correlations, if real,  could give insights into the mechanisms of inhomogeneous broadening and is an area worthy of future investigation.

The $T_1$ relaxation times of around 10\,s are comparable to those seen in saturation recovery methods in measurements of `on-chip' Er:YSO crystals\cite{Probst.2013,Probst.2014}.  
There have also been measurements showing much longer spin-lattice relaxation times, of several hours\cite{Budoyo.2018}, where the lifetime is extended by phonon bottle-necks. 
Like in this work, the experiments in Ref.~\onlinecite{Probst.2013}, with  seconds-scale $T_1$, used coupling to a microwave resonator for readout. This is in contrast to the very long $T_1$ measured with a Josephson bifurcation amplifier\cite{Budoyo.2018}. This suggests the possibility of coupling to the cavity as a relaxation mechanism\cite{Albanese.2020,Bienfait.2016}.

\section{Conclusion}
\label{sec:conclusion}

We have explored the conversion of microwave photons in \Er{}:\YSO{} using a sample free of \Er{167}, at temperatures below 1\,K. Compared to previous measurements of similar systems that used \Er{} in its natural isotopic abundance, we see no optical absorption losses from hyperfine structure in \Er{167}.
We see coherent conversion of microwave photons to optical photons using erbium ions in yttrium orthosilicate, in both the electronic ground $^4I_{15/2} Z_1$ and electronic excited $^4I_{13/2} Y_1$ states. The conversion occurs at the dressed states of the resonator-ion coupled system.
Further, we measured the spin-lattice relaxation time by saturation recovery, finding it to be approximately  10\,s.

\begin{acknowledgments}

We would like to thank Tian Xie and Jake Rochman for useful discussions and practical help setting up the experiments. 
GK wishes to thank Harald Schwefel and Niels Kj\ae{}rgaard for invaluable critical discussions.
This work was supported by the U.S.A.\ Army Research Office (ARO/LPS) (CQTS) Grant No. W911NF1810011, and  the Marsden Fund (Contract No.  UOO1520) of the Royal Society of New Zealand.
\end{acknowledgments}

\bibliography{gkrefs}

\appendix
\section{Simulation of System}
\label{sec:simulate}

The system, comprising an ensemble of erbium ions interacting with microwave and optical fields, was simulated using input-output formalism\cite{Gardiner.1985,QuantumNoise.2004}. 
This generated an equation of the microwave cavity field amplitude, which depends on the interaction between the fields and the ions. From here the dynamics of the ensemble of inhomogeneously broadened ions were modeled using an open quantum systems approach, allowing us to accurately model the interaction between the fields and the ions. Modeling the ions like this allows us to then solve for the intracavity microwave field, and also generates a term proportional to the upconverted optical field.

Using input-output formalism, the equation of motion for the microwave cavity field (with cavity field operator $\hat{b}$) interacting with a microwave transition of the erbium ions is 
\begin{eqnarray}
    \frac{d\hat{b}}{dt}=-i(\wcm\hat{b}+\sum_k g_{\mu}\sigma_{\mu,k})-\frac{\gammc+\gammi}{2}\hat{b}\nonumber\\+\sqrt{\gammc}\hat{b}_{in}+\sqrt{\gammi}\hat{b}_{in,i}
\end{eqnarray}
where $\wcm$ is the microwave cavity central frequency, the sum over index $k$ represents the sum over all the ions, $\sigk{\mu}$ is the atomic transition operator for microwave transition of the $k$th atom, $\gm$ is the coupling strength between the ions and the microwave field, $\gammc$ and $\gammi$ are the coupling and intrinsic losses for the microwave cavity, $\hat{b}_{in}$ represents the input to the cavity through port 1, and $\hat{b}_{in,i}$ represents any other inputs. 

We now make a semi-classical approximation, treating the cavity field as a complex amplitude rather than an operator, $\hat{b}\to\bb$, $\sigk{\mu}$ will also be treated as a complex number corresponding to the coherence between the two levels of the microwave transition, and we set any input not through port 1 to zero, $\bb_{in,i}=0$. Transforming into the frequency domain, this yields,
\begin{equation}
\bb(\delm)=\frac{-i\sum_k\gm\sigk{\mu}}{(\gammc+\gammi)/2-i\delm}+\frac{\sqrt{\gammc}\bb_{in}}{(\gammc+\gammi)/2-i\delm} \label{eq:b_delm}
\end{equation}
where $\delm$ is the detuning between the microwave input field and cavity.
The values for the atomic transition operators depend on the microwave field $\bb$, we now develop a description of the ions which will allow us to find these atomic transition operator terms, and hence the solve for $\bb$. 

For the upconversion process, the ions are modeled as three level atoms, with transitions being driven by the microwave field and the optical pump, and the final $\ket3\to\ket1$ transition corresponding to the output optical photon. The $\ket2$ and $\ket3$ levels are both subject to atomic decoherence and dephasing. The individual ions are modeled using methods described in References~\onlinecite{Barnett.2020} and~\onlinecite{fernandez-gonzalvo.2015}. 
Because there is no cavity enhancement of the optical output field, this field will have a very small amplitude and we assume that this field does not drive the ions. 

The classical values of the atomic transition operators are coherences and are the off-diagonal elements of the atomic density matrix. We can solve for the atomic density matrix using a master equation approach; the time evolution of the density matrix is governed by $\dot{\rho}=\mathcal{L}\rho$, where $\mathcal{L}$ is the Liouvillian accounting for both Hamiltonian evolution and the effects of damping. We can hence solve for the steady state density matrix by solving $\mathcal{L}\rho=0$, accounting for normalization $\sum_i\rho_{ii}=1$

In Equation~\ref{eq:b_delm} the coherences appear as the sum over all of the ions. The ions are all in slightly different environments, which means that they will have slightly different transition frequencies leading to inhomogeneous broadening. Given that there are $\sim 10^{16}$ ions this sum is computationally intractable to perform, and so we approximate it as an integral
\begin{equation}
    \sum_k \gm \sigk{\mu}\approx N_\mu g_\mu \int \sig{\mu}(\delta_2,\delta_3)  G(\delta_2,\delta_3) d\delta_2 d\delta_3 = S_\mu \label{eq:S_mu}
\end{equation}
where $G(\delta_2,\delta_3)$ is a Gaussian distribution representing the inhomogeneous broadening of the ions for both the $\ket2$ and $\ket3$ levels, and the integral is over the entire distribution. This integral is performed numerically, taking into account peaks in $\sig{\mu}(\delta_2,\delta_3)$ due to resonances.

This allows us to solve for Equation~\ref{eq:S_mu} as a function of the microwave field amplitude $\bb$, which in turn allows us to numerically solve for the microwave field amplitude using Equation~\ref{eq:b_delm}. Once we have the intracavity microwave field amplitude, the microwave output field is given by $\bb_{out}=\sqrt{\gammc}\bb$.

Additionally, in solving for $\bb$ we calculate the steady state density matrix, and so calculate all of the steady state coherence terms, not only the coherence in the microwave transition. The ionic $\ket3\to\ket1$ transition corresponds to the generation of an optical output photon, and so the term $\sum_k g_o \sig{13}$ will be proportional to the optical output field, where $g_o$ is the coupling between the transition and the optical field.  

\end{document}